\documentclass[preprint]{revtex4-1}
\usepackage[]{graphicx}

\usepackage{amsmath,units}
\usepackage{mathrsfs}

\graphicspath{{figures/}}

\usepackage[version=3]{mhchem} 

\begin{document}
\title{Active Materials: Biological Benchmarks and Transport Limitations}
\author{Eric R. Dufresne}
\affiliation{Department of Materials, ETH Z\"{u}rich}
\date{March 2019}

\begin{abstract}
 These lecture notes were prepared for the 2018 Summer School on `Active Matter and Non-equilibrium Statistical Physics' at l'\'{E}cole de Physique des Houches.  They survey metabolic activity across a wide range of living organisms, and consider size limitations due to the transport of fuel, waste, and heat for active materials at biomimetic levels of activity.
 
\end{abstract}

\maketitle

\section{Introduction}

Active systems are driven out of thermodynamic equilibrium by microscopic irreversible processes \cite{battle2016}.  
As a class of matter, they present a wide variety of open questions in statistical physics \cite{marchetti2013}, nonlinear dynamics \cite{halatek2018}, and mechanics \cite{needleman2017}.
I would like to focus on \emph{active materials}, a subset of active systems in steady state that possess well-defined continuum properties.
Recent works in cellular and molecular biophysics suggest that activity can drive dramatic changes to the properties of living materials \cite{parry2014,guo2014,humphrey2014,needleman2017} and their fluctuations \cite{lau2003}.

Experimental investigations of active materials pose some challenges.
In order to properly characterize active systems as materials, they need to have a homogeneous composition in steady state at experimentally accessible length scales.
On the other hand, steady state active systems require fluxes of mass and energy to fuel ongoing internal processes.
Since transport rates are limited, this naturally introduces heterogeneity when the physical dimensions of the system are too large.

In this chapter, I will identify some benchmarks for the activity levels of biomimetic active materials and quantify transport constraints on the overall size of homogeneous samples.
More generally, I hope to demonstrate how experimentalists integrate data and some simple theoretical calculations to help design experiments.

\section{Metabolism: the activity of life}

In this section, I briefly describe the biological functions of activity, motivate a simple measure of metabolic activity, and survey its values across a wide range of organisms.

\subsection{Biological Function of Metabolism}

\emph{Metabolism} is the set of chemical processes that underlie homeostasis \cite{alberts2014}, the non-equilibrium steady-state of living systems.
I sort the diverse biological functions of metabolism into two types: physical and information.

The physical functions of metabolism include synthesis, assembly, and motility.  
\emph{Synthetic processes} produce the molecules of life using energy and materials from the environment.  
They include the creation of small molecules, like nucleotides, amino acids, sugars, and lipids, as well as the creation of polymers like nucleic acids, proteins, and carbohydrates.
Living systems use energy to \emph{assemble} structures across length scales.  
At the molecular scale, assembly can be spontaneous or active.  
Examples of spontaneous assembly  include the formation of lipid bilayers and condensation of some membraneless organelles \cite{brangwynne2009}. 
Active assembly takes place over a wider range of length scales.
Chaperones consume energy to assist the folding of single proteins \cite{bukau1998}.
Nucleotide hydrolysis drives the continuous assembly and disassembly of components of the cytoskeleton such as actin filaments and microtubules \cite{pollard2009}.
At the largest scales, the spatial organization of organisms is coordinated through reaction-diffusion schemes \cite{turing1952} and other active mechanisms \cite{gilbert}.
Finally, organisms consume energy to generate \emph{motion} across length scales.  
At the molecular scale, motors such as myosin and kinesin  convert chemical energy into mechanical work \cite{howard2001}. 
At the cellular scale, the cytoskeleton generates forces to change shape or move.
These processes underlie the large scale deformation of tissues, and the movement of organisms.

The information functions of metabolism have two main parts, one for processing and a second for storage.
Cells have elaborate regulatory networks that transduce signals, and process and transmit information \cite{alon2006,lim2014}.
Cells actively transduce physical cues from their environment into chemical information, in the form of the release of small molecules, such as calcium, or the modification of proteins through phosphorylation.
Networks of interactions between  information-carrying proteins consume energy to integrate and process  signals.
Within a cell, information can be transmitted passively by diffusion or with the assistance of active processes, such as the transport of cargo along microtubules.
In higher organisms, neural systems consume energy to transmit information over large distances electrically.
The genetic machinery of life consumes energy to  store and replicate information over evolutionary time scales, \emph{i.e.} millions of years.

It is important to note that, to date, all  investigations of synthetic active matter have focused on activity corresponding to the physical functions of metabolism.
The incorporation of active information functions into synthetic soft materials is an exciting idea, and offers great promise in the long term.  

\subsection{Quantitative approaches to metabolism}

Let's shift gears and think of metabolism chemically.  
To keep it really simple, let's imagine that we are considering a sub-system of a living cell that is  well-mixed.
Then, the state of the system is simply given by the concentration of each of its components, $\phi_i$.
Then, metabolism is captured by the network of reactions between the components, which includes the stoichiometry and rate of each reaction \cite{alon2006}.
From this, the rate of change of the concentration of each species, $d \phi_i / dt$, is determined as well as the rates of change of the associated thermodynamic variables, (\emph{e.g.}  the free-energy   density, $dg/dt$).
In steady state, all of these rates of change are zero, so there must be fluxes of energy and mass at the boundaries to balance the rates of reaction.

In a living system, it's not yet possible to keep track of all the compositions, reactions, and fluxes.  
Instead, people keep track of the total heat or the mass flux of a representative molecule, such as CO$_2$. In the latter case, the results are typically converted to a heat flow rate, $\dot{Q}$, through the enthalpy of glycolysis. 

What can we learn from the heat flow from an active system in steady state?
If the system exchanged only heat with the surroundings, then heat flow would be equivalent to the rate of internal entropy production.
For open systems,  mass fluxes can also carry away entropy, so heat flux is not so easily interpretable. 
Nevertheless, heat flux is readily measured and data is widely available.

\subsection{Survey of metabolic activity in living organisms}

Trans-species surveys of the total metabolic activity, quantified with heat flux, have been popular since the first half of the 20th century \cite{thompson1942,kleiber1947}, as part of a broader quantitative approach to biology called \emph{allometry}.

\begin{figure}
\centering
\includegraphics[width=.6 \textwidth]{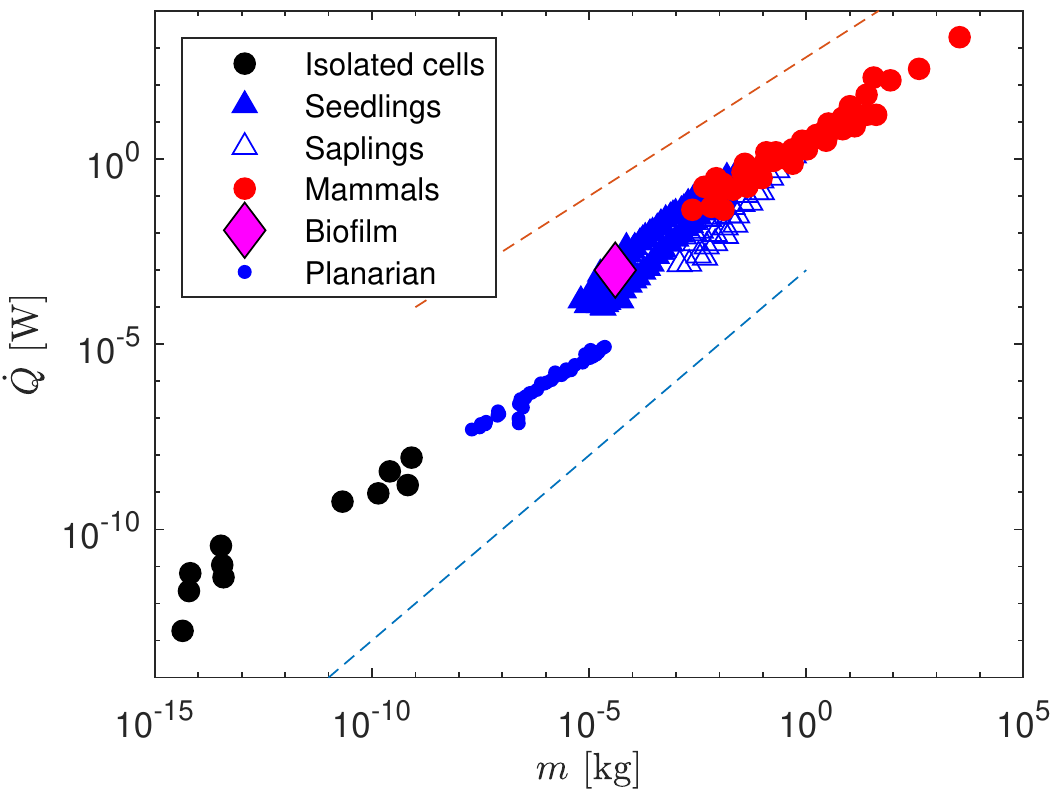}
\caption{\label{fig:Qdot} Metabolic rates  compiled across isolated cells \cite{gillooly2001}, seedlings and saplings \cite{reich2006}, mammals \cite{west2002}, a biofilm \cite{solokhina2017}, and growing planaria \cite{thommen2019}.  Blue and red dashed lines indicate slopes 1 and 3/4 respectively.}
\end{figure}

I have digitized and compiled data on the metabolic activity of living systems from several  original and review papers \cite{west1997,gillooly2001,west2002,reich2006,solokhina2017,thommen2019}.
The total metabolic rate, $\dot{Q}$, is plotted against system mass, $m$, in Fig. \ref{fig:Qdot}.
Over an impressive dynamic range, these diverse systems amazingly appear to follow a simple power-law scaling, $\dot{Q} \sim m^{\nu}$.
Since Kleiber's work on mammals in the 1940's \cite{kleiber1947}, the exponent has been thought to be about 0.75.
Theoretical papers in the late 1990's suggested that this scaling exponent originates from the fractal geometry of the circulatory system \cite{west1997}.
Since then, it has been claimed that the $3/4$-scaling extends down to individual cells \cite{gillooly2001} and even individual organelles \cite{west2002}.
It has recently been observed to hold for individual planaria as they grow \cite{thommen2019}.
On the other hand,  recent work on the growth of seedlings found a linear scaling of the metabolism with mass \cite{reich2006}. 
Nevertheless, this plant data overlaps nicely with the data from mammals \cite{west2002}, as shown in Fig. \ref{fig:Qdot}.
However,  this correspondence only applies to seedlings which are comprised solely of green-tissue.  
The metabolic rate of plants drops below the mammalian trend as they mature into saplings,  where some of the green tissue is replaced with woody tissue.
Presumably, the metabolic rates of mature trees fall substantially further below these values.
A single measurement of the metabolism of a biofilm is similar to seedlings at the same mass \cite{solokhina2017}.

This compelling trend in living systems demands further investigation to reveal its underlying causes.
For now, my aim is not to explain it but to study the basic physics of living materials and design biologically inspired active materials.
To that end, I expect the metabolic density,  $\dot{q}=\dot{Q}/V$, to play an essential role in determining local material properties.  

\begin{figure}
\centering
\includegraphics[width=.6 \textwidth]{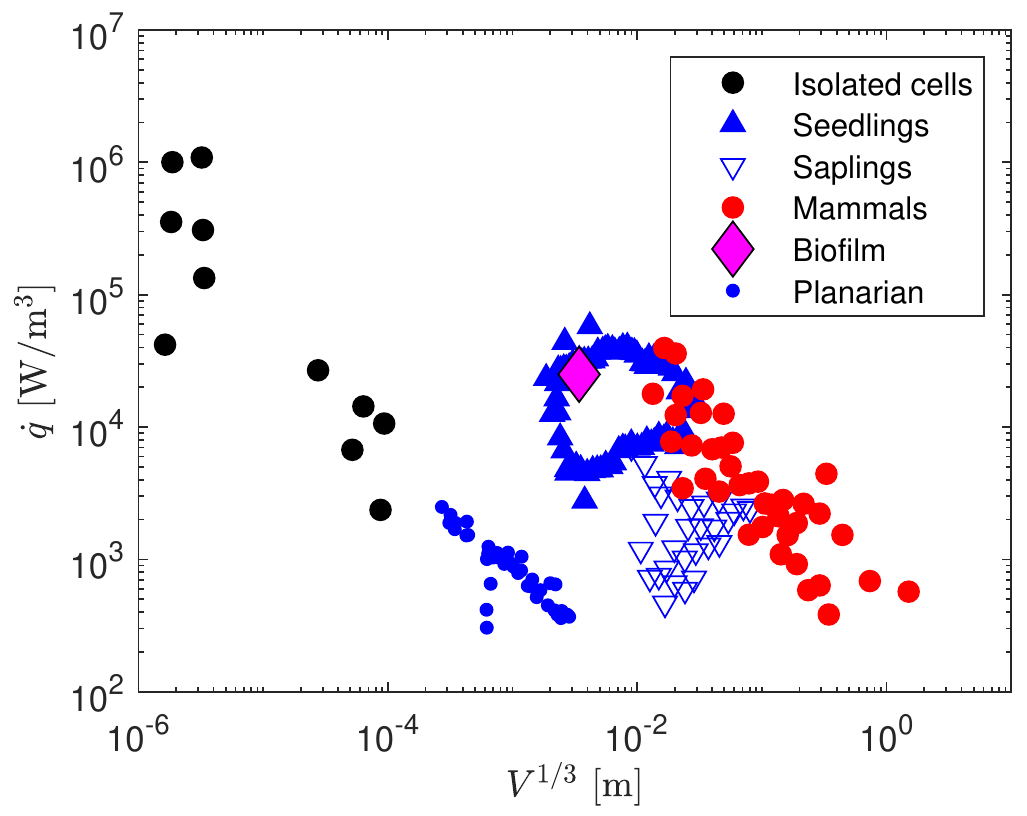}
\caption{\label{fig:qdot} Metabolic densities of  isolated cells \cite{gillooly2001}, seedlings and saplings \cite{reich2006}, mammals \cite{west2002}, a biofilm \cite{solokhina2017}, and growing planaria \cite{thommen2019}. Filled blue triangles indicate the boundary of a dense cloud of data points from \cite{reich2006}. }
\end{figure}

Fig. \ref{fig:qdot} shows the metabolic density calculated from the metabolic rates shown in Fig. \ref{fig:Qdot}, assuming a constant density of $1000~\mathrm{kg/m}^3$.
Here, we see a residual trend with system size, where the smallest systems feature a metabolic density of about $1~\mathrm{MW/m}^3$ and the largest ones generate about $1~\mathrm{kW/m}^3$.
Physical studies of  biomimetic active systems should target this range of values.

\subsection{How large are these metabolic densities?}

Let's put these metabolic densities in familiar context.
The average adult human consumes produces about 100W of heat. 
If we only ate sugar, this would correspond to a diet consisting of 500 g per day.
If we had metabolic densities similar to a bacterium, we would need to eat about 500 kg of sugar per day.
The fabulous power density of bacteria is  comparable to the operational power density of a Li-Ion battery  \cite{battery}.

Now, let's develop some intuition for these values at the molecular scale. 
The free energy liberated by the hydrolysis of ATP, the main energy currency of single molecules, is about $20~k_BT$ \cite{milo2015}.  
At $1~\mathrm{kW/m}^3$, roughly the metabolic density of a human, this corresponds to about one ATP hydrolysis per second in a volume of $(40~\mathrm{nm})^3$.
At $1~\mathrm{MW/m}^3$, roughly the metabolic density of an active bacterium, this corresponds to one ATP hydrolysis per second in a volume of about $(4~\mathrm{nm})^3$.
Bearing in mind that the typical diameter of a protein is about $10~\mathrm{nm}$, this helps us to develop a mental picture of the active processes driving the cytoplasm far from equilibrium.

I now analyze these metabolic densities with a mechanical perspective. 
In a viscous fluid, the volumetric density  of energy dissipation is $\eta \dot{\gamma}^2$.
Here, $\dot{\gamma}$ is the shear rate and $\eta$ is the dynamic viscosity.
Assuming all the metabolic energy is converted into fluid shear, then $ \dot{\gamma}^2=\dot{q}/\eta$.
Thus, metabolic densities of $1~\mathrm{kW/m}^3$ are capable of generating fluid shear rates  around $10^3~\mathrm{s}^{-1}$ in water, and $30~\mathrm{s}^{-1}$ in glycerol, both of which are quite large.
For an elastic solid, the energy density stored in a shear deformation is $G' \gamma^2$, where $G'$ is the shear modulus.  
Thus, the timescale required to achieve a large deformation ($\gamma \approx 1$) is roughly $G'/\dot{q}$.
For muscle ($G' \approx 10~\mathrm{kPa}$) at the average human metabolic density of $1~\mathrm{kW/m}^3$, this is about $10~\mathrm{s}$. 
Since we can bend our elbows in a fraction of a second, it's clear that the peak local metabolic rate in muscle must be much larger than the global average. 

 \section{Transport Limitations}
 
I have established target metabolic densities, based on the range of values found in living systems.
Now, I will determine the constraints imposed by diffusive transport to maintain these metabolic rates.

\subsection{A minimal chemically-driven active system}

\begin{figure}
\centering
\includegraphics[width=.6 \textwidth]{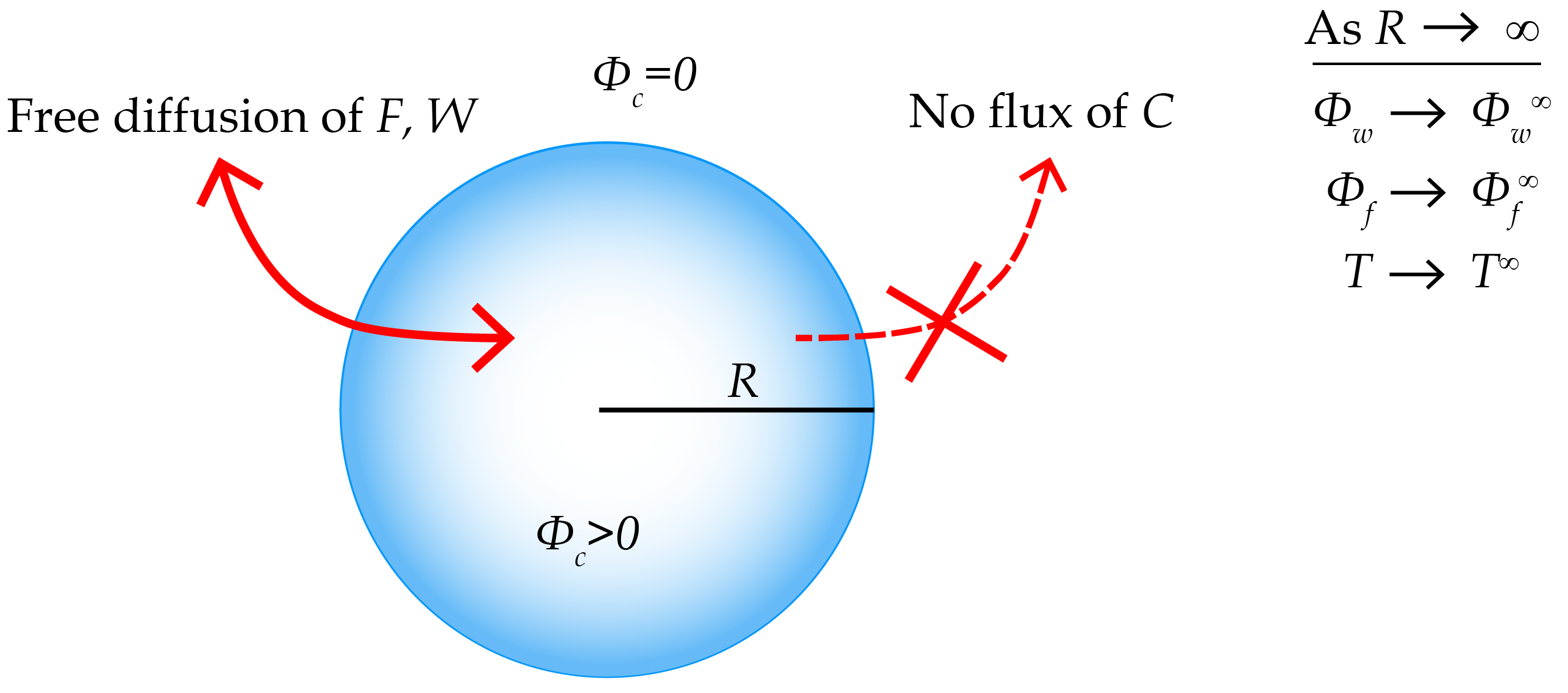}
\caption{\label{fig:minimal} Catalyzed conversion of fuel to waste in a spherical reactive domain.}
\end{figure}

Consider a reactive spherical domain of radius, $R$, at rest with an unbounded fluid domain, schematized in Fig. \ref{fig:minimal}.
This spherical domain perfectly partitions a catalyst, whose  concentration is  $\phi_C$ for $r<R$ and zero for $r>R$, where $r$ is the distance from the center of the domain. 
The surrounding fluid is a reservoir for a chemical fuel, with a concentration that approaches $\phi^{\infty}_F$ far away.
Inside the droplet, the fuel interacts with the catalyst to produce waste, which has a concentration $\phi^\infty_W$ far away.
The temperature far away is $T^\infty$.

We aim to calculate the concentration profiles of fuel and waste, $\phi_F(r)$ and $\phi_W(r)$, as well as the temperature profile, $T(r)$.

\subsection{Kinetics of catalyzed reactions}

We assume that the conversion of fuel into waste by the catalyst proceeds in a simple two-step reaction.
First,  fuel binds to  catalyst with a forward rate $k_f$, reverse rate $k_r$, and equilibrium constant $K_M=k_r/k_f$.
Second, the fuel is converted into waste and released at a rate, $k_{cat}$.
\begin{center}
 \ce{F + C <=>[k_f][k_r] FC ->[k_{cat}] W + C}
\end{center}
If we assume that the concentration of fuel is much higher than the concentration of catalyst, $\phi_F \gg \phi_C$, and the second step of the reaction is slow, $k_{cat} \ll k_r$, then this reaction scheme is equivalent to the classic Michaelis-Menten model of enzyme activity \cite{mm}.


The net reaction rate, $s = d\phi_W/dt = - d\phi_F/dt$, is then given by,
\begin{equation}
s = k_{cat} ~ \phi_C \left( \frac{\phi_F}{K_M+\phi_F} \right).
\end{equation}
At concentrations far above $K_M$, the net reaction rate is limited by the catalyst's internal dynamics  and given by  $k_{cat} \phi_C$.
At concentrations far below $K_M$, the reaction rate is limited by the binding of fuel and catalyst, and is given by \begin{equation}
\label{eq:s}
    s= \frac{k_{cat}}{K_M} \phi_C \phi_F.
\end{equation}

\subsection{Governing Equations of Reaction and Diffusion}
\label{sec:gov}

We assume that the transport of fuel, waste, and catalyst are governed by diffusion.
For species $i$, the diffusive flux is generally, \begin{equation}
    \vec{j}_i = - \sum_j D_{ij}\vec{\nabla} \phi_j,
\end{equation}
where $i,j \in \{F,W,C\}$, and $D_{ij}$ is the diffusion tensor.
The off-diagonal terms of the diffusion tensor create fluxes of one component due to gradients of the others.
This cross-diffusion is a topic of active research for colloids, where it is called diffusiophoresis, and for enzymes, where it is sometimes called `chemotaxis' \cite{anderson1989,dey2014,shi2016,jee2018}.
For simplicity, we assume that there is no cross-diffusion so that this simplifies to \begin{equation}
\label{eq:fick}
    \vec{j}_{F,W} = - D_{F,W}\vec{\nabla} \phi_{F,W}.
\end{equation}

Mass conservation of fuel and waste then requires

\begin{equation}
   \frac{ \partial \phi_{F}}{\partial t} = - \vec{ \nabla} \cdot \vec{j}_{F} - s,
  \end{equation}
  and
  \begin{equation}
   \frac{ \partial \phi_{W}}{\partial t} = - \vec{ \nabla} \cdot \vec{j}_{W} + s.
\end{equation}

In steady state, this simplifies to 
\begin{equation}
    \label{eq:cont}
    s = - \vec{\nabla} \cdot \vec{j}_F =  \vec{\nabla} \cdot \vec{j}_W
\end{equation}
Combining with the expression for the diffusive fluxes, Eq. (\ref{eq:fick}) we find, 
\begin{equation}
\label{eq:diff}
    s = D_F \nabla^2 \phi_F = - D_W \nabla^2 \phi_W.
\end{equation}
Note that in the absence of cross-diffusion, the concentration of the catalyst in steady-state is uniform and equal to $\phi_C$ within the reactive domain.

\subsection{Fuel Profiles}

We will focus on the case where the binding of fuel and catalyst is rate limiting, \emph{i.e} $\phi_F \ll K_M$.
Then, combining Eqs. (\ref{eq:s}) and (\ref{eq:diff}), we arrive at the governing equation for  the fuel concentration \begin{equation}
\label{eq:dh}
    \nabla^2 \phi_F = \kappa^2 \phi_F,
\end{equation}
where the characteristic length scale $\kappa^{-1}$ is given by
\begin{equation}
\label{eq:kappa}
    \kappa^2 = \frac{k_{cat} ~\phi_C}{K_M D_F}.
\end{equation}
Within the reactive domain ($r<R$), this equation is formally identical to the Debye-H\"{u}ckel equation for linear screened electrostatics.  
There, $\kappa^{-1}$ is the `screening length' which determines the distance over which the electrostatic potential decays.
Here, we will see that $\kappa^{-1}$ plays a similar role for the fuel concentration, and I  refer to it as the \emph{penetration length}.
Outside the reactive domain ($r>R$), where $\phi_C=0$, this reduces to the LaPlace equation of unscreened electrostatics.
Thus, in that region we expect the depletion of fuel  to fall off like $1/r$.

The appropriate boundary conditions to solve Eq. (\ref{eq:dh}) are as follows.  
First, we require finite $\phi_F$ everywhere and that it approaches $\phi_F^\infty$ at $r \gg R$.
Next, we require continuity of $\phi_F$ and $\vec{j}_F$ at $r=R$, despite the discontinuity in $\phi_C$.

With a bit of math, we can now find the concentration profile within the reactive domain,
\begin{equation}
    \phi_F(r<R) = \frac{\phi_F^\infty}{\cosh (\kappa R)} \frac{\sinh (\kappa r)}{\kappa r},
\end{equation}
and outside the reactive domain,
\begin{equation}
    \phi_F(r>R) = \phi_F^\infty \left( 1 + \frac{\tanh( \kappa R) - \kappa R}{\kappa r} \right).
\end{equation}

\begin{figure}
\centering
\includegraphics[width=.95 \textwidth]{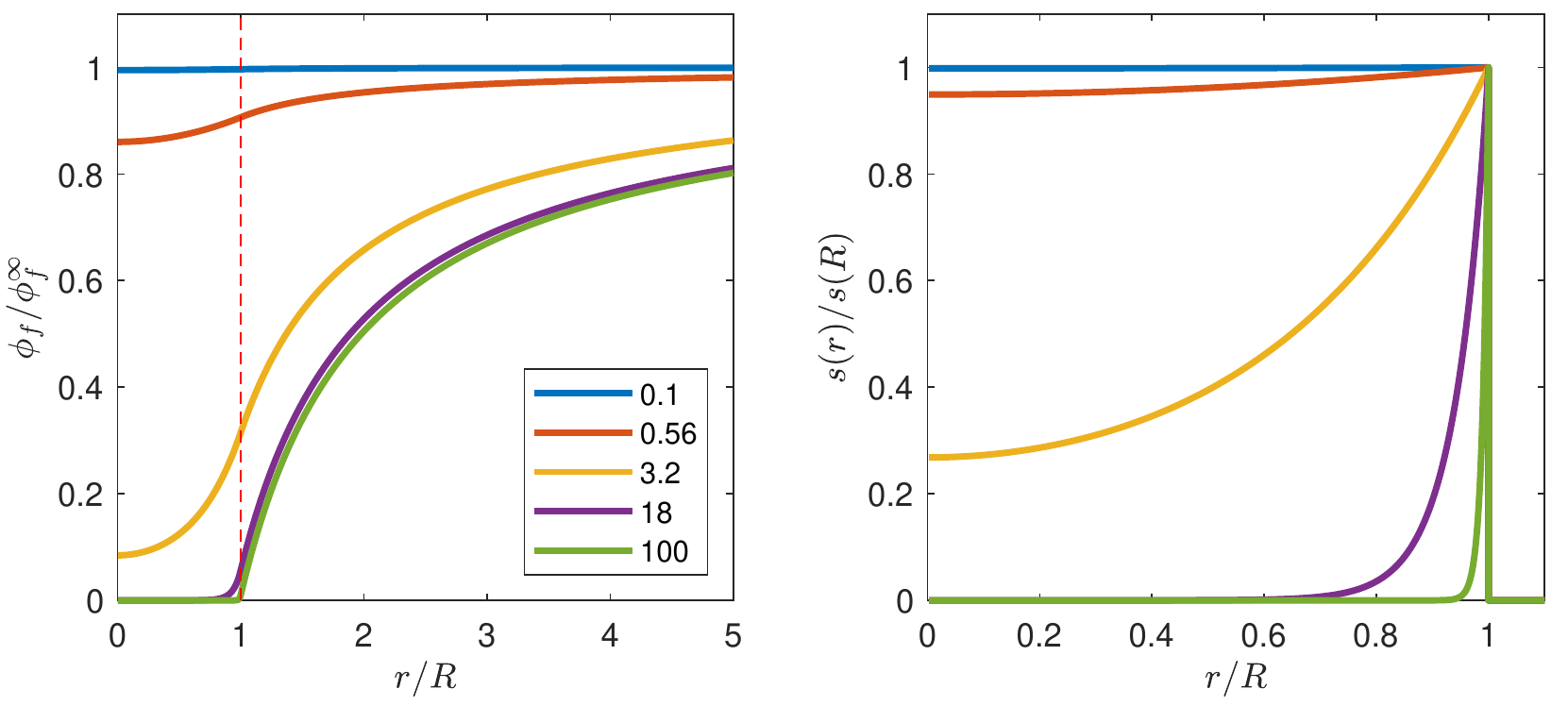}
\caption{\label{fig:fuel_profile} Fuel profile \emph{(left)} and reaction rate \emph{(right)} for $\kappa R = [0.1,~0.56,~3.2,~18,~100]$.  }
\end{figure}

The fuel profiles are governed by the dimensionless parameter $\kappa R$, the ratio of the domain size to the penetration length.
Fuel profiles are shown for a range of $\kappa R$ values from 0.1 to 100 in Fig. \ref{fig:fuel_profile} \emph{(left)}.
At $\kappa R=0.1$, the fuel concentration is essentially uniform.
Here, the conversion of fuel to waste is slow compared to its diffusion.
At $\kappa R = 100$, the concentration of fuel is strongly perturbed.  
The fuel is almost fully depleted within the droplet, and the fuel concentration is also strongly suppressed in the vicinity of the droplet.

The profile of the reaction rate, $s(r)$, is plotted for a range of $\kappa R$ values in Fig. \ref{fig:fuel_profile} \emph{(right)}.
Here, the reaction profiles are normalized by the rate at the boundary $s(R)$.
For $\kappa R \ll 1$, the reaction rate is uniform within the reactive domain.
For $\kappa R \gg 1$, the reaction rate is limited to a small layer at the surface of the domain, whose thickness is given by the penetration length.



It is now clear that $\kappa R$ determines the spatial distribution of the activity.
What are realistic values for $\kappa$?
From Eq. (\ref{eq:kappa}), it appears that we may need to specify the molecular details of the interaction of the fuel and  catalyst, because it involves the product $k_{cat} \phi_C/K_M$.
However,  this combination of variables also appears in Eq.(\ref{eq:s}).  
 In the limit $\kappa R \ll 1$, where the reaction rate in the catalytic domain is uniform, this  becomes $s=(k_{cat}/K_M) \phi_C \phi_F^{\infty}$.
Finally, we can infer the reaction rate from the metabolic density through the reaction's change in enthalpy, $s \approx \dot{q}/\Delta H$.
Combining these relationships, Eq. (\ref{eq:kappa})
 can be conveniently expressed in terms of macroscopic quantities
\begin{equation}
    \kappa^2 \approx \frac{\dot{q}}{\Delta H D_F \phi^\infty_f}.
\end{equation}

Now, we can calculate numerical values of the penetration length for a specific reaction scheme.
Let's consider a common source of energy in living systems: glycolysis.  
Here, glucose is broken down to release energy according to \ce{C6H12O6 + 6O2 ->  6CO2 + 6H2O}, with a release of enthalpy of about $\Delta H = 3~\textrm{MJ}/\textrm{mol~\ce{C6H12O6}}$.  
The diffusion coefficient of glucose in water is about $D_F=7 \times 10^{-10}~\mathrm{m}^2/\mathrm{s}$.
Finally, we pick a fuel concentration $\phi_F^\infty = 5~\mathrm{mM} = 5~\mathrm{mol}/\mathrm{m}^3$, a typical glucose concentration in human blood.
Therefore, as the metabolic density increases from $10^3$ to $10^6~\mathrm{W}/\mathrm{m}^3$, the penetration depth drops from $3~\mathrm{mm}$ to $0.1~\mathrm{mm}$.

\subsection{Transport of Waste and Heat}

The above calculation suggests that  active materials with biologically relevant metabolic densities can be homogeneous when the overall system size is small, $\kappa R \ll 1$.  
However, our calculation  has only considered fuel transport.  
The removal of waste and heat are also essential.
Now, we consider their transport in the limit $\kappa R \ll 1$.  


We already derived the governing equation for the waste concentration in Section \ref{sec:gov}, \begin{equation}
\nabla^2 \phi_W = - \frac{s}{D_W}.
\end{equation}
Following a similar line of reasoning, one can derive the governing equation for the temperature, \begin{equation}
    \nabla^2 T = -\frac{s \Delta H}{\rho c \alpha}.
\end{equation}
Here, $\rho$ is the density, $c$ is the specific heat, and $\alpha$ is the thermal diffusivity.
The product $\rho c \alpha$ is commonly known as the thermal conductivity, and is about $0.6~\mathrm{ W/K~ m}$ in water.
These two equations have the same form.
In contrast to the governing equation for the fuel, Eq. (\ref{eq:diff}), they
have no characteristic length scale.
Their solutions are given by
\begin{equation}
\frac{\phi_W(r)-\phi_W^\infty}{\Delta \phi_W^o}=\frac{T(r)-T^\infty}{\Delta T^o} = 
            \left\{
                \begin{array}{ll}
                  1-\frac{1}{3} \left( \frac{r}{R}\right)^3, ~r<R\\
                  \frac{2}{3} \left( \frac{R}{r} \right), ~ r>R
                \end{array}
              \right..
\end{equation}
As expected, this solution shows that the temperature and waste concentrations are elevated in the reactive domain.
The excess waste concentration at the center of the domain is \begin{equation}
    \Delta \phi_W^o \equiv \phi_W(0)-\phi_W^\infty = \frac{sR^2}{2D_W} \approx \frac{\dot{q}}{2 \Delta H }\frac{R^2}{D_W},
\end{equation} and the temperature increase at the center of the domain is \begin{equation}
    \Delta T^o \equiv T(0)-T^\infty = \frac{s \Delta H}{\rho c} \frac{R^2}{2 \alpha}
    \approx \frac{\dot{q}}{\rho c} \frac{R^2}{2 \alpha}.
\end{equation}
Note that these values  scale with the square of the radius of the reactive domain.  

How large are these values at physiological metabolic rates?
In the previous section, we determined the maximum size for a reactive domain that maintains a uniform reaction rate.
At $10^3~\mathrm{W/m^3}$, this was about 1 mm.
At $10^6~\mathrm{W/m^3}$, this was about 30 $\mathrm{\mu m}$. 
In both of these limiting scenarios, $\Delta \phi^o_W \approx 0.5~\mathrm{mM}$ and $\Delta T^o \approx 1~\mathrm{mK}$.
Here, we have used 
$\Delta H = 0.5~\mathrm{MJ}/\mathrm{mol~ \ce{CO2}}$ and  $D_W=2\times10^{-9}~\mathrm{m}^2/\mathrm{s}$.
While this temperature difference is a tiny fraction of the ambient temperature, the excess concentration of \ce{CO2} is closer to its concentration in physiological conditions, about 30 mM. 
This reflects a general tendency for mass transport  to be rate-limiting,  as thermal diffusion coefficients are usually  larger than mass diffusion coefficients.


\section{Conclusions}

We surveyed the metabolic activity of a wide range of living systems, from single cells to large mammals and found that the average metabolic density varies from about $1~\mathrm{kW}/\mathrm{m}^3$ to $1~\mathrm{MW}/\mathrm{m}^3$.
With an eye toward making synthetic materials that mimick these living systems, we considered the constraints that the transport of fuel, waste, and heat put on system size.
We found that homogeneous physiological metabolic densities can be readily maintained in small-scale systems.
At bacterial metabolic densities, this corresponds roughly to systems below 10 $\mu$m in radius.  
At metabolic densities comparable to an average human cell, this corresponds to systems below about 1 mm.
Intriguingly, these cutoff length scales are about 10 times greater than the respective sizes of bacteria and human cells.
Since these dimensions are much larger than molecular scales, a continuum description of the properties of active materials at physiological metabolic densities seems reasonable.
Furthermore, these scales are readily accessible using modern experimental tools.

It is important to emphasize some of the limitations of the transport calculations in these notes.  
First, I have ignored any type of convection.  
External convection (\emph{e.g.} a steady flow to remove waste and provide more fuel) could relax constraints due to heat and waste transport, but will not change constraints due to the reaction-diffusion of the fuel inside the reactive domain.
Convection within the reactive domain would relax the constraints due to the fuel.  
This could be achieved by  spontaneous motion within the reactive domain \cite{keber2014}, or by `brute force' through the introduction of a biomimetic circulatory system. 
Naturally, transport limitations play an important role as the driving force for the development of  circulatory systems in living systems \cite{truskey2004}.
Second, I have ignored cross-diffusion \cite{anderson1989}.
Mobility of catalyst in fuel-gradients could potentially destabilize the system and drive heterogeneity. 

Armed with these biological benchmarks and  an understanding of the constraints of diffusive transport, we can now begin to design experiments to probe the properties of synthetic materials at biomimetic levels of activity.



Acknowledgements:  Andrea Testa, Robert W. Style, Yingjie Xiang, Ernst Hafen, Raghuveer Parthasarathy, Nicolas Bain, and the students, lecturers, and organizers of the Summer School on `Active Matter and Non-equilibrium Statistical Physics' at l'\'{E}cole de Physique des Houches.

%

\end{document}